\documentclass{article}

\usepackage[final]{neurips_2021}

\usepackage[utf8]{inputenc} 
\usepackage[T1]{fontenc}    
\usepackage{hyperref}       
\usepackage{url}            
\usepackage{booktabs}       
\usepackage{amsfonts}       
\usepackage{nicefrac}       
\usepackage{microtype}      
\usepackage{xcolor}         
\usepackage{natbib}
\usepackage{caption}
\usepackage{subcaption}
\usepackage{graphicx}
\graphicspath{'images/'}
\usepackage{float}

\usepackage{enumitem}

\title{Semi-Supervised GCN for learning Molecular Structure-Activity Relationships}

\author{%
 Alessio Ragno\thanks{Correspondence to \href{mailto:ragno@diag.uniroma1.it}{ragno@diag.uniroma1.it}}\  $^1$, Dylan Savoia$^1$, Roberto Capobianco$^{1,2}$\\\\
  $^1$Dept. of Computer, Control and Management \\Engineering ``Antonio Ruberti'', 
  Sapienza University of Rome \\$^2$Sony AI\\
}

\begin{document}

\maketitle

\begin{abstract}
    Since the introduction of artificial intelligence in medicinal chemistry, the necessity has emerged to analyse how molecular property variation is modulated by either single atoms or chemical groups. In this paper, we propose to train graph-to-graph neural network using semi-supervised learning for attributing structure-property relationships. As initial case studies we apply the method to solubility and molecular acidity while checking its consistency in comparison with known experimental chemical data. As final goal, our approach could represent a valuable tool to deal with problems such as activity cliffs, lead optimization and de-novo drug design.
\end{abstract}

\section{Introduction}

Drug discovery is a process that consists in searching, developing and optimizing molecules in order to disclose new lead compounds for future new drugs as medicaments for specific diseases. 

One big issue in drug discovery consists in experimentally evaluating the molecules, a part of the drug development flow that is time-consuming and expensive. 

In order to compensate this problem, medicinal chemists generally use computational approaches to derive models for forecasting molecules' bio-activity and selecting the ones associated to higher predicted values, for example, quantitative structure-activity relationships (QSAR) consist in the application of machine learning for predicting the molecules' activity from their structure. 

In this paper we propose a method to study these relationships calculating atom attributions through their graph representation. It is, in fact, very important for the chemist to individuate the most relevant substructures in order to exploit them for the development of new molecules. Once a QSAR model is built, we propose to train a graph-to-graph network (GGN) to attribute the molecules' atoms for their properties.



\section{Related Work}

While, with classic machine learning, molecules are represented using fingerprints and descriptors, after the introduction of graph convolutional networks (GCN) \citep{gcn} it became possible to represent their structure using graphs. This allows us to learn over molecules taking advantage of both chemical and structural information. 
When studying molecules' properties, it is not only interesting to be able to predict them using different estimators, but, it is important to understand how they are modulated by the different functional groups.

In order to study these structure-activity relationships, researches have been exploiting explainable AI (XAI) methods for interpreting the GCN predictions. This allows to visualize the atoms that are considered ``mostly responsible'' for modulating the activity. Some approaches exploit integrated gradients (IG) \citep{Taly2017} to generate input-attributions on the atoms \citep{ig_mols}. Although XAI methods do a good job in attributing specific atoms, it is difficult for them to refer to functional groups. Moreover, many approaches \citep{GNNExplainer, PGExplainer, GraphMASK} involve the perturbation of graphs by means of atom removal, masking the graphs or
edge cutting, but they do not take into account the fact that, in chemistry, removing an atom or a bond from a molecule might generate an impossible structure to realize. Other techniques \citep{GraphLIME} focus, instead, on the perturbation of the node features, but again, this could cause the creation of non-sense values.

\section{Proposed Approach}

In this paper, we propose to train a GGN model, using a semi-supervised approach \citep{gcn}, which is able to give scores to atoms and functional groups of a molecule, depending on their influence on a QSAR model's prediction. To train the model, we need to set some ground-truth output values for the atoms: we modify the molecules by adding or substituting atoms and we evaluate the activity of the generated molecules, finally, we compare them with the original one. We then train the GGN model on the modified molecules, predicting the score of the nodes corresponding only to the added atoms. In this way, the resulting model is able to give scores to every atom of the molecule, depending on its influence on the property. 

With this idea, we build a dataset by generating new molecules via atom addition and set the ground-truth scores: given a molecule $m$, with predicted activity $a$, and a second molecule $m'$, with predicted activity $a'$, derived from $m$ through an addition of a group $g$, the ground-truth scores for the atoms of $g$ are defined as $attr(g_i) = \frac{a' - a}{|g|}$. We train the GGN to approximate $attr(g_i)$, with $g_i \in g$.

\begin{figure}[t!]
  \centering
  \includegraphics[width=.5\textwidth]{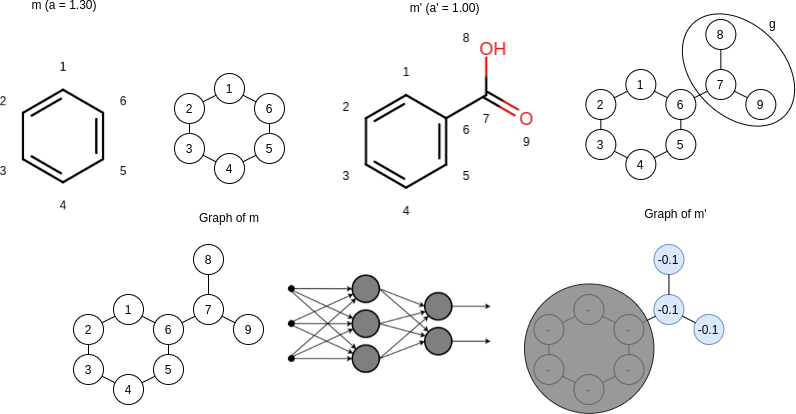}
  \caption{Ground-truth score generation example with benzene and benzoic acid.}
  \label{fig:generation}
\end{figure}

For instance, given a QSAR model and molecule $m$ with a predicted activity $a = 1.30$ (benzene in \figurename~\ref{fig:generation}) we can add a group $g$ of 3 atoms (carboxyl group in \figurename~\ref{fig:generation}) generating a second molecule $m'$ (benzoic acid in \figurename~\ref{fig:generation}). Assuming the evaluation using the QSAR model of $m'$ is $a'=1.00$, we know that the addition led an activity decrement of $1.00 - 1.30 = -0.30$, therefore we can assume that the contribution of the added atoms would be $\frac{-0.30}{3} = -0.10$.

In order to make modifications to the molecules, we need a set of functional groups to add. \citet{FGGen} proposes an algorithm to extract functional groups from a set of molecule and applied it to $\approx~$483,000 molecules with activity below 10 $\mu$M on any target retrieved from ChEMBL database \citep{ChEMBL}. The resulting database consists in a set of 768 groups that are present in at least 10 ChEMBL molecules, provided in pseudo-SMILES notation.

\begin{table}[b!]
\centering
\begin{tabular}{@{}ccccc@{}}
\toprule
\textbf{Molecule} & \textbf{IG} & \textbf{DeepLIFT} & \textbf{GGN} & \textbf{logS} \\ \midrule
hexane            & -0.1166     & -0.1196           & -1.8828               & -9.1150       \\
pentane           & -0.0932     & -0.0956           & -1.8627               & -7.5483       \\
1-hexanol         & -0.0657     & -0.0711           & -0.5852               & -2.8134       \\
1-pentanol        & -0.0404     & -0.0460           & -0.5696               & -1.3863       \\
3-pentanol        & -0.0096     & -0.0163           & -0.1369               & -0.5447       \\\midrule
\textbf{Correlation with logS}      & 0.96                     &     0.95                               & \textbf{0.98}                                    & 1.0                             \\\bottomrule
\end{tabular}%
\caption{Molecular scores obtained with IG, DeepLIFT and the proposed approach compared with the logS of the molecules}
\label{tab:pentane}
\vspace{1em}
\end{table}
\section{Experimental Results}
We evaluate our method on two datasets:
\setlist{nolistsep}
\begin{itemize}[noitemsep]
    \item ESOL Water Solubility Dataset \citep{Delaney2004}, which contains more than 1k molecules with the associated logarithm of the solubility expressed in molar concentration (logS);
    \item A set of 160k molecules retrieved from ChEMBL \citep{ChEMBL} with the associated ChemAxon predicted pKa.
\end{itemize}




We train two GCN-QSAR models on the datasets and use the functional groups database to generate the dataset for the GGN model with the ground-truth attribution scores. Finally for each initial dataset, a GGN model is trained.
The following subsections analyse the attributions to the GCN-QSAR predictions obtained with some molecules for both the tasks comparing them with two XAI methods: IG \citep{Taly2017} and DeepLIFT \citep{DeepLIFT}. 
The attributions are analysed through a mask of colors on the atoms on a scale from blue to red, where blue is negative and red is positive. The opacity of the color indicates the intensity of the scores.
To produce a quantitative evaluation we sum up all the atom scores for each molecule: we expect molecules with a higher value of the property in exam to have a higher molecular scores. Finally we analyse the correlation between molecular scores and their properties.

\subsection{Water Solubility}
Water solubility is very important in medicinal chemistry, since most of the drugs must be present in the form of solution at the site of absorption. Moreover, solubility works as a benchmark since we know from chemical theory that the presence of specific atoms or patterns modulates significantly the property.

As first analysis the attributions for 5 different molecules external to the dataset with incremental solubility are compared:  hexane, pentane, 1-hexanol, 1-pentanol and 3-pentanol.

\begin{table}[b]
\centering
\begin{tabular}{@{}crrrr@{}}
\toprule
\textbf{Molecule}          & \multicolumn{1}{c}{\textbf{IG}} & \multicolumn{1}{c}{\textbf{DeepLIFT}} & \multicolumn{1}{c}{\textbf{GGN}} & \multicolumn{1}{c}{\textbf{pKa}} \\ \midrule
pentanoic acid             & 0.4884                          & 0.9821                                & 1.6650                                    & 5.01                             \\
4-chloropentanoic acid     & 0.1337                          & 0.0765                                & 0.5025                                    & 4.27                             \\
3-chloropentanoic acid     & 0.1610                          & 0.2180                                & 0.5006                                    & 4.24                             \\
2-chloropentanoic acid     & 0.0253                          & -0.0779                               & 0.4777                                    & 3.96                             \\\midrule
\textbf{Correlation with pKa}   & \textbf{0.98}                          & 0.41                               & \textbf{0.98}                                    & 1.0                                \\ \bottomrule

\end{tabular}
\caption{Molecular scores obtained with IG, DeepLIFT and the proposed approach compared with the pKa of the first pKa case study molecules}
\label{tab:pentanoic}
\end{table}

Chemical theory and experimental values for solubility of the molecules in exam suggest that the hydroxyl (OH$^-$) group generally increases the solubility if introduced in a molecule and the more the it is at the center of the carbon chain, the more the solubility increases.
Hexane and pentane are only composed of a chain of carbons, while 1-hexanol and 1-pentanol are generated adding an hydroxyl group to the formers, respectively, this makes them more soluble than their original molecules and in fact the hydroxyl group receives positive scores from all the methods (\figurename~\ref{fig:pentane}, rows 3 and 4). Using the same reasoning, the 3-pentanol, is more soluble than the previous molecules, as its hydroxyl group is placed at the center of the chain. Also in this case, the attribution methods report coherent values with the chemical theory (\figurename~\ref{fig:pentane}, row 4). The molecular scores obtained with the different methods report values in good agreement with the expectation, in fact the molecular scores have a high positive correlation with the logS (\tablename~\ref{tab:pentane}).



\subsection{pKa}

The acidity of the molecules is a property which  influences the solubility of drugs and their pharmacodynamics and pharmacokinetics (ADMET) properties. The strength of an acid is generally measured through the acid dissociation constant, $K_a$, or its linearized version pKa.

\begin{figure}[t]
     \centering
     \begin{subfigure}[t]{0.4\textwidth}
         \centering
             \includegraphics[width=\textwidth]{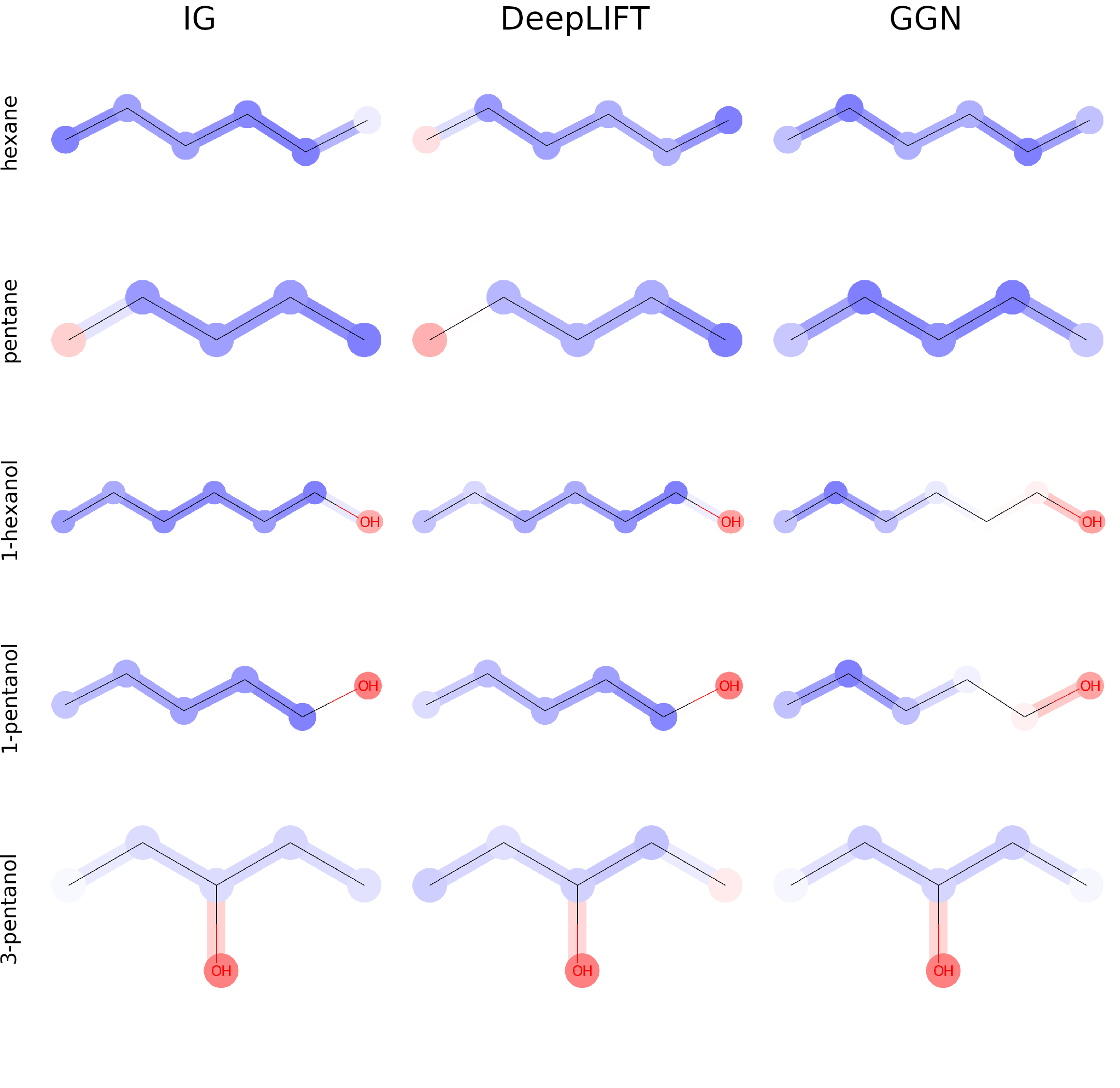}
    \caption{Solubility attribution comparison varying hydroxyl group position}
    \label{fig:pentane}
     \end{subfigure}
     \hfill
     \begin{subfigure}[t]{0.4\textwidth}
          \centering
         \includegraphics[width=\textwidth]{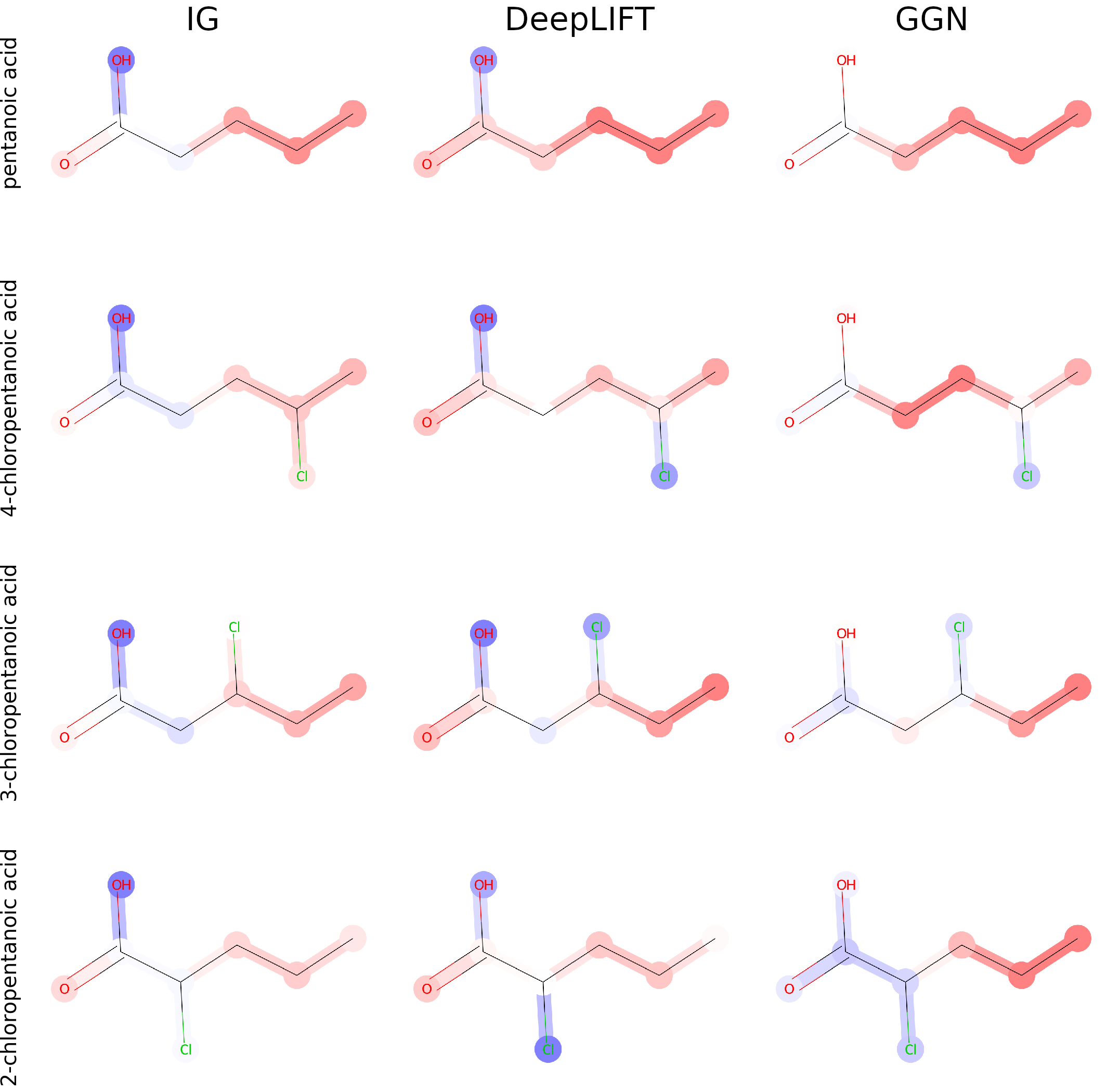}
         \caption{pKa attribution comparison varying the chlorine position}
        \label{fig:pentanoic}
     \end{subfigure}
     \caption{Molecular attributions compared between IG, DeepLIFT and the proposed approach}
\end{figure}

We take into exam five molecules: pentanoic acid, 4-chloropentanoic acid, 3-chloropentanoic acid and 2-chloropentanoic acid. In this case study, the molecules in exam have a negative trend of pKa, due to the position of the chlorine along the molecules: the closer to the carboxyl group, the more pKa decreases (more acid). 
The carboxyl group in the pentanoic acid is correctly attributed with the lowest score since it is the group that determines the molecule's acidity (\figurename~\ref{fig:pentanoic}, row 1). Moreover, our method is the only one which correctly identifies the carboxyl group as a whole. For the other molecules, only our method and DeepLIFT are able to attribute the chlorine as an element which influences negatively on the pKa (\figurename~\ref{fig:pentanoic}, rows 2, 3 and 4).
Although DeepLIFT seems to perform similarly to our approach, correlation between the molecular scores and pKa values show that the our methods's results are more coherent with the experimental values (\tablename~\ref{tab:pentanoic}). 

\section{Conclusions}
\label{conclusions}

Neural networks in drug design may help reducing the costs of lead discovery and, with GCNs, it is possible to exploit both chemical and structure information of the molecules thanks to the graph representation. We only performed the experiments using the vanilla GCN layer as the main goal was to provide a general method. Other variants, such as Graph Attention Layers, could be used as well, both for the QSAR model and the GGN.

XAI proved to be a key tool for developing future models in drug design. Although novel techniques are able to individuate the most relevant atoms, it is still difficult for them to focus on functional groups. The proposed approach managed to overcome this issue and we think that it could be extended in order to optimize molecules' activities/properties by adding or substituting atoms in a rational way.

As a drawback, the proposed method consists in building a GGN network. This approach proved to be more ``chemically oriented'' but brings the overload of training an additional model.
\pagebreak
\medskip
\bibliographystyle{plainnat}
\bibliography{bibliography} 

\end{document}